\documentclass[preprint]{JHEP3} 

\JHEPspecialurl{http://jhep.sissa.it/JOURNAL/JHEP3.tar.gz}

\usepackage{epsfig,multicol}

\def\beq{\begin{equation}}

\def\eeq{\end{equation}}
\def\eq#1{{Eq.~(\ref{#1})}}

\newcommand{\bea}{\begin{eqnarray}}

\newcommand{\eea}{\end{eqnarray}}

\parskip=\itemsep               

\newcommand{\beqar}[1]{\begin{eqnarray}\label{#1}}

\newcommand{\eeqar}{\end{eqnarray}}

\def\thefootnote{\fnsymbol{footnote}} 

\title{
{\Large  \bf The Yin and Yang of High Energy Chromodynamics: Scattering in Black and White.}}
\author{A. Kovner and M. Lublinsky \\
Physics Department, University of Connecticut, \\ 2152 Hillside
Road, Storrs, CT 06269-3046, USA}

\abstract{We further discuss the QCD Reggeon field theory (RFT) as it emerges from the  JIMWLK/KLWMIJ evolution equation and beyond.
We give an explicit expression for the calculation of scattering amplitude in terms of the eigenstates of the RFT Hamiltonian. We point out that the spectrum of RFT is doubly degenerate, the degeneracy being related to the spontaneous breaking of the Dense-Dilute Duality symmetry of RFT. The degeneracy is between the "almost white" states (the Yang sector) which contain a small number of gluons, and "almost black" states (the Yin sector). The excitations above the Yang vacuum have natural interpretation in terms of gluons. Analogously the excitations above the Yin vacuum  have natural interpretation as "holes" in the black disk - points at which an incoming gluon does not scatter with unit probability. We discuss in detail the spectrum of the "parton model approximation" to the KLWMIJ evolution introduced in our previous paper,  and prove that it is explicitly selfdual. This allows us to find explicitly the counterpart  hole states in this approximation. We also present an argument to the effect that the end point of the evolution for any physical state cannot be a "grey disk" but must necessarily be the "black disk" Yin vacuum state.
Finally, we suggest an approximation scheme for including the Pomeron loop contribution to the evolution which requires only the solution of the JIMWLK/KLWMIJ Hamiltonian.
}

\begin{document}



\def\thefootnote{\arabic{footnote}}

\section{Introduction}

In this paper we continue our investigation of the correspondence between the Reggeon Field Theory (RFT)\cite{Gribov, Amati, Baker} and the modern incarnation of the ideas of gluon saturation \cite{GLR,MUQI,Mueller,mv} as encoded in the JIMWLK-KLWMIJ functional evolution equation \cite{balitsky,Kovchegov,JIMWLK,cgc,kl}.
In \cite{KLreggeon} we have pointed out that the JIMWLK-KLWMIJ Hamiltonian can be directly interpreted as the Hamiltonian of RFT as arising from the first principles QCD calculation. The basic quantum field theoretical degree of freedom in this Hamiltonian are the unitary matrix $R(x)$, which is the single gluon scattering matrix. In this respect this RFT differs from the original Gribov's template considered in \cite{Amati,Baker} and also recently in \cite{BF} which deals directly with scattering of color neutral physical objects. It is rather more conceptually related to Lipatov's approach \cite{BFKL} in which the basic entity is the color nonsinglet reggeized gluon \cite{LipatovRgluon}.

We have shown that the reggeized gluon as well as other colored reggeon like states arise as the eigenstates of the RFT Hamiltonian in the partonic type approximation which preserves the number of $s$-channel gluons throughout the energy evolution. In the same approximation we have found the BFKL Pomeron as well as other Pomeron like and Odderon eigenstates.
Thus the parton like approximation to the RFT reproduces the known perturbative QCD exchanges
as well as contains other exchanges which only appear if multiple scattering of a single projectile gluon with the target is allowed. These additional exchanges grow faster with energy but their coupling is suppressed by powers of  $\alpha_s$\footnote{The parton like approximation is designed for the situation when the projectile is small, while the target is far from being black. Therefore, just like the original BFKL approximation it is not applicable at asymptotic energies, and in fact it violates the unitarity of the scattering amplitude.}.

The spectrum of the full KLWMIJ (or equivalently JIMWLK) hamiltonian is not known apart from two eigenstates which both have zero eigenvalue.
We will call these two states "white" or Yang, and "black" or Yin. The white state is simply the QCD vacuum state - the field strength in this state is zero. The scattering matrix of any projectile on this state is unity and it does not evolve with energy. The "black" state corresponds to the black disc limit. The probability density to find any gluon field configuration in this state is independent of configuration, and the scattering matrix of any projectile on it vanishes.
These two states are related to each other by the Dense-Dilute Duality (DDD) transformation discussed in \cite{something}. The DDD transformation was shown in \cite{something, balitsky1} to be an exact symmetry of the full RFT hamiltonian which should include the Pomeron loop effects\cite{shoshi1,IMM,IT,kl,kl4,MSW,LL3,levin}. Although the JIMWLK-KLWMIJ hamiltonian is not self dual, the remnant of this symmetry survives on the zero eigenvalue subspace. 
The full selfdual RFT Hamitonian (which we will refer to as $H^{RFT}$) is not known yet, although a number of selfdual extensions of the JIMWLK evolution has been proposed and derived in some approximations \cite{kl1,SMITH, Balitsky05} (see however \cite{kl5}).

In the present paper our goal is different. Rather than study specific approximation schemes or specific models, we aim to discuss some general properties of the spectrum of RFT, specifically those properties which follow from the exact selfduality which the full Hamiltonian of RFT must possess \cite{something,balitsky1}.

The structure of this paper is the following. In Section 2 we recall the basic equations of the high energy evolution in the JIMWLK-KLWMIJ framework. We also give a derivation of the general form of the evolution in terms of $n$- gluon emission amplitudes. This section sets the stage and notations for the subsequent discussion. The relation of the high energy evolution kernel to the multigluon emission amplitudes to our knowledge has not been discussed in the literature so far. It is useful for the understanding of the general structure of the evolution but is not essential for the rest of the discussion in this paper.

In Section 3 we show that the Hilbert space of the full (and as yet unknown) RFT Hamiltonian splits into two orthogonal subspaces which we dub the Yin and the Yang Hilbert spaces. The Yang states are those close to the white vacuum. They contain a small number of gluons and have $S$-matrix close to unity. The dual image  of this space is the Yin space. These are states close to the black vacuum. To characterize these states we find it natural to introduce the concept of a "hole". The hole at a transverse coordinate $x$ is a state  close to black disk, such that a gluon that scatters on it has unit probability to scatter unless it hits precisely at the point $x$, in which case it has a finite probability to propagate through the target without scattering. This concept naturally arises in the context of JIMWLK evolution, where the single gluon scattering matrix $S$ acts as a hole creation operator.

We illustrate the Yin-Yang structure within the partonic approximation to the KLWMIJ evolution introduced in \cite{KLreggeon}. We show that this approximation is itself selfdual, albeit under duality transformation which is somewhat modified relative to that discussed in \cite{something}. Using the results of \cite{KLreggeon} we find explicit eigenfunctions in the Yin sector with up to two holes.

In Section 4 we discuss how the eigenfunctions of the evolution Hamiltonian enter the calculation of physical scattering amplitudes. To this end we recall the restrictions on allowed states of RFT imposed by their interpretation from the point of view of scattering of physical gluons. We show that these restrictions are such that any physical state must have a nonvanishing projection on the Yin vacuum state. We also write down the expression for the physical $s$ - matrix in terms of the eigenfunctions of $H^{RFT}$. Considering carefully the restriction on physical states we argue that  asymptotically at large energies all physical states become black. Thus if an additional "grey" vacuum state exists it decouples from the evolution of physical states.

Finally in Section 5 we conclude by discussing an approximation scheme to $H^{RFT}$ which includes Pomeron loops but requires 
only the knowledge of the solutions of $H^{JIMWLK}$.

\section{The high energy evolution and the gluon emission amplitudes}

The process we are interested in is the scattering of a highly energetic left moving projectile consisting of gluons on a hadronic target. We are working in the lightcone gauge natural to the projectile wave function, $A^-=0$. 
In this gauge the high energy scattering matrix of a single  gluon projectile at
 transverse position $x$ on the target is given by 
the eikonal factor  \footnote{In our convention
the variable $x^-$ is rescaled to run from 0 to 1.}
\begin{eqnarray}\label{S}
S(x;0,x^-)\,\,=\,\,{\cal P}\,\exp\left\{i\int_{0}^{x^-} dy^-\,T^a\,\alpha^a_T(x,y^-)\right\}\,;
\,\,\,\,\,\,\,\,\,\,\,\,\,\,\,\,\,\,\,S(x)\,\equiv\,S(x;0,1)\,.
\end{eqnarray}
where $T^a_{bc}=if^{abc}$ is the generator of the $SU(N)$ group in the adjoint representation and $x_i$ - the transverse coordinate. 

The field $\alpha_T$ is the large $A^+$ component created by the target color charge density. It
obeys the classical equation of motion  and is determined by the color charge density of the target $\rho_T(x)$ via \cite{first,second,JIMWLK,cgc} 
\beq\label{alpha}
\alpha^a_T(x,x^-)T^a\,\,=g^2\,\,{1\over \partial^2}(x-y)\,
\left\{S^\dagger(y;0,x^-)\,\,\rho^{a}_T(y,x^-)\,T^a\,\,S(y;0,x^-)\right\}\,.
\eeq
 
For a composite projectile which has some distribution of gluons in its wave function
 the eikonal factor can be written in the form analogous to 
$S(x)$, see \cite{something}
\begin{equation}
\Sigma^P[\alpha_T]\,\,=\,\,\int D\rho_P\,\,W^P[\rho_P]\,\,
\,\exp\left\{i\int_{0}^{1} dy^-\int d^2x\,\rho_P^a(x,y^-)\,\alpha_T^a(x,y^-)\right\}\,.\label{s}
\end{equation}
Here the subscripts $P$ and $T$ refer to the projectile and target resectively.
The quantity $\rho_P(x_i)$ is the color charge density in the projectile wave function at a given transverse position, while $W^P[\rho]$ can be thought of as the weight functional
which determines the probability density to find a given configuration of color charges in the projectile. For a single gluon $\rho^a(x_i)=T^a\delta^2(x_i-x^0_i)$, and eq.(\ref{s}) reduces to eq.(\ref{S}). 

The total $S$-matrix of the scattering process at a given rapidity $Y$ is given by
\begin{equation}
{\cal S}(Y)\,=\,\int\, D\alpha_T^{a}\,\, W^T_{Y_0}[\alpha_T(x,x^-)]\,\,\Sigma^P_{Y-Y_0}[\alpha_T(x,x^-)]\,.
\label{ss}
\end{equation}
In \eq{ss} we have restored the rapidity variable and have chosen the frame where the target has rapidity $Y_0$ while the projectile carries the rest of the total rapidity $Y$.
Lorentz invariance  requires ${\cal S}$ to be independent of $Y_0$ \cite{LL1,LL2}. 

The high energy evolution is given by the following general expression\footnote{We have adopted the conventions in which the spectrum of the Hamiltonian $H^{RFT}$ is positive, and not negative as is standard in the recent literature, see for example  \cite{KLreggeon}. This allows us to think about the system described by $H^{RFT}$ as of a standard stable quantum mechanical system.}:
\begin{equation}
-\frac{d}{d\,Y}\,{\cal S}\,=\,\int\, D\alpha_T^{a}\,\, W^T_{Y_0}[\alpha_T(x,x^-)]\,\,\,
H^{RFT}\left[\alpha_T,\frac{\delta}{\delta\,\alpha_T}\right]\,\,\,
\Sigma^P_{Y-Y_0}[\alpha_T(x,x^-)]\,.
\label{hee}
\end{equation}
Here $H^{RFT}$ stands for a generic kernel of high energy evolution, which can be viewed as acting
either to the right or to the left, as it should be Hermitian:
\begin{equation}
-{\partial\over\partial Y}\,\Sigma^P\,\,=\,\,H^{RFT}\left[\alpha_T,\,{\delta\over\delta\alpha_T}\right]\,\,\Sigma^P[\alpha_T]\,;
\ \ \ \ \ \ \ \ \ 
-{\partial\over\partial Y}\,W^T\,\,=\,\,
H^{RFT}\left[\alpha_T,\,{\delta\over\delta\alpha_T}\right]\,\,W^T[\alpha_T]\,.
\label{dsigma}
\end{equation} 

A complete quantitative treatment of the evolution is yet unavailable. It has been shown however in \cite{something} that in order for the total $S$-matrix to be Lorentz invariant and symmetric between the projectile and the target, the evolution kernel $H^{RFT}$ must be self dual. That is it has to be invariant under the Dense-Dilute Duality (DDD) transformation
\begin{equation}\label{duality}
\alpha^a(x,x^-)\,\rightarrow \,i\,{\delta\over\delta\rho^a(x,x^-)}\,,\ \ \ \ \ \ \ \ \ \ \ \ \, {\delta\over\delta\alpha^a(x,x^-)}
\,\rightarrow \,-\,i\,\rho^a(x,x^-)\,.
\end{equation}

The derivation of the kernel is available in two limits - the high density limit where the color charge density in the wave function is assumed to  be $O(1/\alpha)$ and the low density limit where the number of gluons in the wave function is assumed to be $O(1)$.
Both these limits break the symmetry between the target and the projectile \cite{kl5}.

In the limit when the color charge density of the target is parametrically large ($\rho^a=O(1/\alpha_s)$) the kernel is given by
 the JIMWLK expression \cite{JIMWLK,cgc}
\begin{equation}\label{jimwlk}
H^{JIMWLK}\,=\,{\alpha_s\over 2\pi^2}\,\int d^2z \,\bar Q^a_i(z)\,\bar Q^a_i(z)\,=\,
{\alpha_s\over 2\pi^2}\,\int d^2z\, \tilde {\bar Q}^a_i(z)\, \tilde {\bar Q}^a_i(z)
\end{equation}
where the hermitian "amplitudes" $\bar Q^a_i(z)$ are defined as
\begin{equation}
\bar Q^a_i(z)=\int d^2x{(x-z)_i\over (x-z)^2}[S^{ab}(z)-S^{ab}(x)]\bar J^b_R(x);\ \ \ \ \ \ 
\tilde {\bar Q}^a_i(z)=\int d^2x{(x-z)_i\over (x-z)^2}[S^{ba}(z)-S^{ba}(x)]\bar J^b_L(x)\,.
\end{equation}
The generator of the right color rotations $J_R$ and the generator of the left color rotations $J_L$ are defined as\cite{kl1}
\begin{equation}
\bar J_R^a(x)=-{\rm tr} \left\{S(x)T^{a}{\delta\over \delta S^\dagger(x)}\right\}, \ \ \ \ \bar J_L^a(x)=
-{\rm tr} \left\{T^{a}S(x){\delta\over \delta S^\dagger(x)}\right\}, \ \ \ \  \ \ \ \ \ \ \ 
\bar J_L^a(x)\,\,=\,\,[S(x)\,\bar J_R(x)]^a\,.
\end{equation}
In the representation where the JIMWLK Hamiltonian acts on functionals of $\alpha(x^-,x)$, the rotation operators are 
\begin{equation}
\bar J_R^a(x)={\delta\over\delta\alpha(x,x^-=1)}; \ \ \ \ \ \ \ \ \ \ \ \ \ \ \bar J_L^a(x)={\delta\over\delta\alpha(x,x^-=0)}\,.
\label{barjays}
\end{equation}
The JIMWLK Hamiltonian is usually written in a form different from eq.(\ref{jimwlk}), see for example \cite{kl4}. It is however straightforward to check that eq.(\ref{jimwlk}) is indeed equivalent to the standard JIMWLK kernel by using the unitarity of matrix $S$.

The KLWMIJ Hamiltonian (the dilute target limit) can be written in a very similar manner
\begin{equation}\label{klwmij}
H^{KLWMIJ}\,=\,{\alpha_s\over 2\pi^2}\,
\int d^2z\, Q^a_i(z)\,Q^a_i(z)\,=\,{\alpha_s\over 2\pi^2}\,\int d^2z\, \tilde Q^a_i(z)\,\tilde Q^a_i(z)
\end{equation}
where the hermitian "amplitudes" $\tilde Q^a_i(z)$ are defined as
\begin{equation}
 Q^a_i(z)\,=\,\int d^2x{(x-z)_i\over (x-z)^2}\,[R^{ab}(z)\,-\,R^{ab}(x)]\, J^b_R(x);\ \ \ \ \ \ \ \tilde Q^a_i(z)=\int d^2x{(x-z)_i\over (x-z)^2}[R^{ba}(z)-R^{ba}(x)]J^b_L(x)
\label{qus}
\end{equation}
where $R^{ab}(x)$ is the color charge shift operator
\begin{equation}
R(x)\,=\,{\cal P}\exp\left\{\int dx^-\, {\delta\over\delta\rho^a(x,x^-)}\,T^a\right\}\,.
\label{alpha1}
\end{equation}
The generator of the right color rotations $J_R$ and the generator of the left color rotations $J_L$ are defined as
\begin{equation}
J_R^a(x)=-{\rm tr} \left\{R(x)T^{a}{\delta\over \delta R^\dagger(x)}\right\}, \ \ \ \  J_L^a(x)=
-{\rm tr} \left\{T^{a}R(x){\delta\over \delta R^\dagger(x)}\right\}, \ \ \ \  \ \ \ \ \ \ \ 
J_L^a(x)\,\,=\,\,[R(x)\, J_R(x)]^a,
\end{equation}
or in the representation analogous to eq.(\ref{barjays}) 
\begin{equation}
J_R^a(x)=\rho^a(x,x^-=1); \ \ \ \ \ \ \ \ \ \ \ \ \ \ \ J_L^a(x)=\rho^a(x,x^-=0)\,.
\label{jays}
\end{equation}
The JIMWLK and KLWMIJ Hamiltonians are clearly related by the DDD transformation.

In the rest of this section we present a framework for derivation of these formulae which gives a  more general perspective on the high energy evolution.

Consider a wave function of a hadron at certain rapidity $|\Psi\rangle$. We are interested in calculating expectation values of a variety of observables in this wave function. In particular we want to calculate average of the $S$-matrix - an operator which depends only on the charge density operators $\hat\rho^a(x)$. In general we will consider any operator that depends on $\hat\rho$: $\hat O[\rho(x)]$. As we have discussed in detail in \cite{kl, kl4} such a calculation can be represented as a path integral over $\rho(x,x^-)$, where $x^-$ is the ordering variable.
\begin{equation}
\langle\Psi|\,\hat O[\hat\rho(x)]\,|\Psi\rangle\,=\,\int D\rho(x,x^-)\, W[\rho(x,x^-]\,\,O[\rho(x,x^-)]
\end{equation}
with a weight functional $W$, where the function $O[\rho(x,x^-)]$ is defined by expanding the operator $\hat O$ in power series in $\hat \rho$ and assigning the value of $x^-$ to each factor of $\hat \rho$ in the order it appears in the product. The exact value of $x^-$ does not matter as the correlators $\rho(x^-_1)...\rho(x^-_n)$ depend only on the ordering of $x^-_i$ but not on their exact values. 

It is convenient to represent this expectation value in a different way. Define the charge density shift operator
\begin{equation}
\hat R\left[{\delta\over\delta\beta}\right]\,=\,{\cal P}
\exp\left\{\int dx^-\int d^2x \,\hat\rho^a(x)\,{\delta\over\delta\beta^a(x^-,x)}\right\}
\end{equation}
where the path ordering is along the $x^-$ direction and the integration is over the entire $x^-$ axis. We can then write
\begin{equation}
\hat O[\hat\rho(x)]\,=\,\hat R\left[{\delta\over\delta\beta}\right]\,\,O[\beta(x,x^-)]|_{\beta=0}\,;\ \ \ \ \ \ \ \ \ \ \ \ \ \ \ 
\langle\Psi|\,\hat O[\hat\rho(x)]\,|\Psi\rangle\,=\,\Gamma\left[{\delta\over\delta\beta}\right]\,\,O[\beta(x,x^-)]|_{\beta=0}
\end{equation}
where
\begin{equation}\label{gammabeta}
\Gamma\left[{\delta\over\delta\beta}\right]\,=\,
\langle\Psi|\,\hat R\left[{\delta\over\delta\beta}\right]\,|\Psi\rangle\,=\,
\int D\rho(x,x^-)\, W[\rho(x,x^-]\,\hat R\left[{\delta\over\delta\beta}\right]\,.
\end{equation}
Therefore as far as the averaging over the wave function is concerned one always has to calculate the average of the operator $\hat R$ 
rather than that of an arbitrary operator $\hat O$.
Curiously, this expression can be written in the form very similar to that for the $S$-matrix derived in \cite{kl4}. The wave function $\Psi$ in the Fock basis is simply a collection of creation operators acting on the vacuum
\begin{equation}
|\Psi\rangle\,=\,\Psi[a^\dagger]\,|0\rangle\,.
\end{equation}
The operator $\hat R$ when acting on the gluon creation operator at transverse coordinate $x$ rotates it by the "$c$-number" matrix
\begin{equation}
R_\beta(x)\,=\,{\cal P}\exp\left\{\int dx^- \,T^a\,{\delta\over\delta\beta^a(x^-,x)}\right\}\,.
\end{equation}
Thus
\begin{equation}
\Gamma\left[{\delta\over\delta\beta}\right]\,=\,\langle\Psi[a]\,|\,\Psi[R_\beta\,a^\dagger]\rangle \,.
\end{equation}

Consider now the same hadron boosted to a slightly higher energy so that the rapidity of all the partons changes by $\Delta Y$. The wave function of the hadron changes. This change can be represented as the action of the unitary "cloud operator" 
${\cal C}[\hat \rho, a_\Delta, a^\dagger_\Delta]$ where by $a_\Delta$ and $a^\dagger_\Delta$ we cumulatively denote the annihilation and creation operators of gluons at rapidities in the new part of phase space opened up by the boost. 
\begin{equation}
|\Psi\rangle\,\rightarrow \,{\cal C}|\Psi\rangle\,.
\end{equation}
The explicit form of the operator $\cal C$ is known in the low density regime (see \cite{zakopane} for a detailed discussion). Our derivation here is general,  not restricted to the dilute limit and does not require this knowledge.
The state $|\Psi\rangle$ does not contain the soft gluons created by $a^\dagger_\Delta$:
\begin{equation}\label{a}
a_\Delta\,|\Psi\rangle\,=\,0.
\end{equation}
The color charge density changes due to the charge of the extra emitted gluons by
\beq
\Delta\rho^a(x)\,=\,a^\dagger_\Delta(x)\, T^a\,a_\Delta(x)\,.
\eeq
The density shift operator becomes
\begin{equation}
\hat R_\Delta\left[{\delta\over\delta\beta}\right]\,=\,
{\cal P}\exp\left\{\int dx^-\,d^2x\, [\rho^a(x)\,+\,\Delta\rho^a(x)]\,{\delta\over\delta\beta^a(x^-,x)}\right\}\,.
\end{equation}

The expectation value of the same observable in the boosted wave function is now given by
\begin{equation}
\langle\Psi|{\cal C}^\dagger\,\hat O[\rho^a(x)\,+\,\Delta\rho^a(x)]\,{\cal C}|\Psi\rangle\,=\,
\Gamma_\Delta\left[{\delta\over\delta\beta}\right]
\,\, \hat O[\beta(x,x^-)]|_{\beta=0}
\end{equation}
\beq
\Gamma_\Delta\left[{\delta\over\delta\beta}\right]\,=\,
\langle\Psi|{\cal C}^\dagger\,\hat R_\Delta\left[{\delta\over\delta\beta}\right]\,{\cal C}|\Psi\rangle\,.
\eeq
Note that by virtue of eq.(\ref{a})
\beq
\hat R_\Delta\left[{\delta\over\delta\beta}\right]\,|\Psi\rangle\,=\,\hat R\left[{\delta\over\delta\beta}\right]\,
|\Psi\rangle=\Psi[R_\beta\,a^\dagger]|0\rangle \,.
\eeq
The operator $\Gamma$ therefore changes as a result of the evolution as
\begin{equation}\label{evolve}
\Delta\Gamma\,=\,\Gamma_\Delta\,-\,\Gamma\,
=\,
\langle 0|\Psi[a]\left\{\,\,\ {\cal C}^\dagger\,\hat R_\Delta \,{\cal C} \,\hat R^\dagger_\Delta\,\,-\,\,1 \,\,\right\}
 \Psi[R\,a^\dagger]|0\rangle \,.
\end{equation}
The operator 
$\hat R_\Delta\,{\cal C}\, \hat R^\dagger_\Delta$ has all the charge density operators and field operators rotated by $R_\beta$ 
relative to the operator $\cal C$
\begin{equation}
\hat R_\Delta\, {\cal C} [\hat\rho(x),\,\,a_\Delta(x),\,\ a^\dagger_\Delta(x)]\,\hat R^\dagger_\Delta\,=\,
{\cal C} [R_\beta(x)\hat\rho,\,\,R_\beta(x)a_\Delta(x),\,\,R_\beta(x)a^\dagger_\Delta(x)]\,.
\end{equation}
Also, as shown in \cite{kl4}, every operator $\hat \rho$ inside the averaging, when acting to the left/right, 
becomes the operator of the left/right rotation $J_{L,R}(R_\beta)$ (\ref{jays}), plus the ordering of the operators gets reversed. Let us 
formally expand the operator $\cal C$ in the Fock basis of the operators $a_\Delta$
\begin{equation}
{\cal C}[\hat\rho(x),\,\,a_\Delta(x),\,\,a^\dagger_\Delta(x)]|0_\Delta\rangle\,=\,
\sum_n\,c_n[\hat\rho]\,\,a_\Delta^{a_1\,\dagger}(x_1)\,\cdots\, a_\Delta^{a_n\,\dagger}(x_n)|0_\Delta\rangle
\end{equation}
where $|0_\Delta\rangle$ denotes the Fock vacuum of the operators $a_\Delta$.
Here the index $n$ denotes the number of soft gluons as well as their color indices and transverse coordinates, so that less concisely
\beq
c_n[\hat\rho]\,\equiv\, c^{a_1,a_2...,a_n}_{x_1,x_2,..., x_n}[\hat\rho]\; \ \ \ \ \ \ \ \ \ \ \ \ \ \ \ \ \ 
c_n^\dagger[\hat\rho]\,\equiv\, c^{b_1,b_2...,b_n}_{x_1,x_2,..., x_n}[\hat\rho]\,.
\eeq
Then we have
\begin{equation}\label{exp}
\Delta\Gamma\,=\,\left\{\sum_n\,c^\dagger_n[J_L]\,R_{\beta}^{b_1 a_1}(x_1)\,\cdots\,R_{\beta}^{b_n a_n}(x_n)\,c_n[J_R]\,\,-\,\,1\right\}\, \Gamma
\end{equation}
where the color indices of $R_\beta$ are contracted with those of the coefficient functions $c$.
In the dilute limit only one gluon is emitted per one step of the evolution \cite{kl} and
the KLWMIJ equation is derived  such that only the term $n=1$ is kept in \eq{exp}. In general however it is clear that multiple gluons are emitted, even though the explicit form of $\cal C$ is not known in a general situation.

We can obtain another representation of $\Delta\Gamma$ concentrating on Hermitian 
observables $\hat O$ only. For these observables 
\begin{equation}
2\,\Gamma\left[{\delta\over\delta\beta}\right]\,=\,
\langle\Psi|\,\hat R\left[{\delta\over\delta\beta}\right]\,\,+\,\,\hat R^\dagger\left[{\delta\over\delta\beta}\right]\,|\Psi\rangle\,.
\end{equation}
For any unitary $A$ one has $2-A-A^\dagger=(1-A^\dagger)(1-A)$ and we can use this "optical theorem" to rewrite
eq.(\ref{evolve})
\begin{equation}
\Delta\Gamma\,=\,-\,
{1\over 2}\,\langle 0|\Psi[a]\,\,\left(1\,-\,{\cal C}^\dagger\,\hat R_\Delta {\cal C}\, \hat R^\dagger_\Delta\right)\,
\left(1\,-\,\hat R_\Delta \,{\cal C}^\dagger\,\hat R^\dagger_\Delta \,{\cal C}\right)\,\,\Psi[R_\beta\,a^\dagger]|0\rangle \,
=\,-\,{1\over 2}\,\sum_n
Q^\dagger_n\,Q_n\,\Gamma\,.
\end{equation}
The second equality is obtained by inserting the resolution of identity on the soft gluon Fock space, 1\,=\,$\sum_n |n_\Delta\rangle\,\langle n_\Delta|$
with the $n$ soft gluon state defined as
\beq
|n_\Delta\rangle\,=\,a_\Delta^{a_1\,\dagger}(x_1)\,\cdots\, a_\Delta^{a_n\,\dagger}(x_n)\,|0_\Delta\rangle
\eeq
and
\begin{equation}
Q_n[R_\beta]\,=  \,\langle n_\Delta|\,1\,-\,{\cal C}^\dagger[J_R,a_\Delta,a^\dagger_\Delta]\,\,
{\cal C}[J_L, R_\beta\, a_\Delta, R_\beta\,
a^\dagger_\Delta ]\, |0_\Delta\rangle\,.
\end{equation}
Each $Q_n$ depends on transverse coordinates and has a set of $n$ Lorentz and color indices which we do not indicate explicitly.

Since $\Gamma[{\delta\over\delta\beta}]$ by definition is the Fourier transform of $W[\rho]$ with $\rho$ and $\delta/\delta\beta$ being conjugates, eq.(\ref{gammabeta}) , the evolution of $W[\rho]$ is obtained from the evolution of $\Gamma[{\delta\over\delta\beta}]$ simply by replacing $\beta\rightarrow \rho$ in the evolution kernel.
The infinitesimal form of the evolution is obtained by expanding $Q_n$ to the lowest order in the evolution parameter $\Delta Y$. 
Thus the "weight function" $W[\rho]$ evolves according to
\beq
-{\partial\over\partial Y}W[\rho]=H^{RFT}W[\rho]
\eeq
with 
\beq\label{hrft}
H^{RFT}\,=\,{d\over d \Delta Y}\,{1\over 2}\,\sum_n\,Q^\dagger_n[R]\,Q_n[R]|_{\Delta Y=0}
\eeq
where the unitary operators $R(x)$ is defined in terms of the derivatives with respect to the color charge density eq.(\ref{alpha1}). 

To construct $H^{RFT}$ explicitly  we need to know the form of the cloud operator $\cal C$. However there are several properties of the evolution which are independent of explicit form of $\cal C$. First, since the evolution operator is the sum of squares, it is clearly positive definite\footnote{The derivative in the definition eq.(\ref{hrft}) does not affect the conclusion about positive definiteness, since the product $Q^\dagger_n Q_n$ vanishes at $\Delta Y =0$ while it is positive for any nonzero $\Delta Y$.}. Second, this form of the evolution explicitly preserves the signature symmetry $R\rightarrow R^T$ as defined in \cite{KLreggeon} irrespective of the explicit form of $\cal C$, since hermitian conjugation  of $Q_n$ simply takes $R$ into its transposed.

The meaning of $Q_n$ is the amplitude for creation of $n$ gluons as a result of the evolution to higher rapidity. 
In the KLWMIJ limit (\ref{klwmij}) only one gluon is emitted in one step in rapidity, thus the single gluon emission amplitude determines the evolution as in eq.(\ref{qus}). In this case the index $n$ denotes the color, the transverse polarization and the transverse position of the emitted gluon. The form of the emission amplitude eq.(\ref{qus}) is precisely the same as that of the inclusive single gluon emission amplitude for the scattering on the target described by the scattering matrix $R$ \cite{bkw}.  Beyond the KLWMIJ limit more than one gluon is emitted in one step of the evolution. For example in order to obtain the JIMWLK limit an arbitrary number of gluons would have to be emitted in one step of the evolution. This is consistent with the cloud operator being a Bogoliubov type operator \cite{inprep}.

\section{The Yin and the Yang}
Our main point of interest is how to understand the evolution of the physical $S$-matrix eq.(\ref{ss}) in terms of 
solutions of $H^{RFT}$. The natural strategy for calculating the $S$-matrix is the following.
First find eigenfunctions of the RFT Hamiltonian, 
\begin{equation}\label{eigen1}
H^{RFT}\,|\Psi_i^{l_i}\rangle\,=\,\omega_i\,|\Psi_i^{l_i}\rangle
\end{equation}
where the index $l_i$ labels degenerate eigenstates corresponding to the eigenvalue $\omega_i$.
The evolution of each eigenfunction is given by
\begin{equation}
|\Psi_i^{l_i}\rangle_Y\,=\,e^{-\,\omega_i \,Y}\,|\Psi_i^{l_i}\rangle\,.
\end{equation}
Since $H^{RFT}$ must be hermitian, its eigenfunctions must form a complete orthonormal basis.
Next expand $\Sigma^P$ at initial rapidity in this basis 
\beq\label{bas}
\Sigma^P\,=\,\sum_{i,l_i}\,\gamma_{i,l_i}^P\,|\Psi_i^{l_i}\rangle
\eeq
and similarly expand $W^T$:
\beq\label{basw}
W^T\,=\,\sum_{i,l_i}\,\ \gamma^{T\,*}_{i,l_i}\,\langle \Psi_i^{l_i}|
\eeq
The total cross section at rapidity $Y$ is 
\beq\label{over}
{\cal S}(Y)\,=\,\sum_{i,l_i,l'_i}\,\gamma^{P}_{i,l_i}\, \epsilon_i^{l_i\,l'_i}\,\gamma^{T\,*}_{i,l'_i}\,\,e^{-\,\omega_i\,Y}\,
\eeq
where  $\epsilon_i^{l_i\,l'_i}$ is the overlap matrix between degenerate eigenstates at a given value of $\omega_i$.

In this discussion as well as in the rest of this section the scalar product and the norm of the eigenfunctions 
$|\Psi_i\rangle$ are defined in the standard way
\beq\label{hnorm}
\langle\Psi_i|\Psi_{i'}\rangle\,=\,\int D\rho \,\Psi^*_i[\rho]\,\Psi_{i'}[\rho]\,.
\eeq
 Note that $\Sigma$ and $W$ are not necessarily normalized to unity with respect to this norm, thus  
$\sum_i\,\gamma_i^{P\,*}\,\gamma_i^P\,\ne\, 1$ etc. The general properties of $W$ and $\Sigma$ have been discussed to some extent in \cite{KLreggeon}. We will return to this question in the next section.

The question we are addressing in this section is how the DDD symmetry affects the spectrum of RFT. 
Since RFT is formulated as a regular quantum field theory with a hermitian Hamiltonian, quite generally any symmetry of this Hamiltonian must be represented in the spectrum in either Wigner or Goldstone mode. That is it is either the symmetry of the spectrum, or is spontaneously broken in the vacuum. In the latter case the vacuum has to be degenerate and the Hilbert space has to split into direct sum of two Hilbert spaces each one built over the appropriate vacuum state. 

Is DDD spontaneously broken or not in RFT? A little thought is required to see that the answer is affirmative. The simple argument is the following. The QCD vacuum state does not contain gluons and is Lorentz invariant - therefore it does not contain gluons in any frame. This state therefore is a vacuum of the RFT - the S-matrix of scattering it on any hadron is always unity and does not depend on rapidity 
($\omega_0=0$). This is the white "Yang" vacuum of RFT. On the other hand the DDD transformation as shown in \cite{something} does not leave this state invariant, but rather transforms it into the completely black state - the "Yin" vacuum. Common sense of course tells us that the totally black state is also invariant under the high energy evolution, since the probability of any hadron to scatter on it is strictly equal to unity. Beyond the simple common sense the degeneracy of the Yin and Yang vacua is ensured by the fact that DDD is the symmetry of the RFT Hamiltonian. 

\subsection{The two vacua}

The two vacuum structure is present already in the KLWMIJ and JIMWLK Hamiltonians, even though the DDD transformation is not a symmetry of each one of them separately. Consider for example the KLWMIJ Hamiltonian. It has been derived in the limit of a dilute system and thus must by definition contain the Yang vacuum. This is indeed the case.
The eigenstates satisfy the Schroedinger type equation
\begin{equation}
H^{KLWMIJ}\,\Psi[\rho]\,=\,\omega\,\Psi[\rho]\,.
\end{equation}
The form of $H^{KLWMIJ}$ as sum of squares ensures its positive definiteness. Thus the only zero eigenstates are those annihilated by all $Q_i$ in eq.(\ref{qus}). 
We find useful in the following to use both the "configuration space basis" and the "momentum space basis". The first one is the basis of eigenstates of (mutually commuting) operators $\rho^a(x,x^-)$:
\begin{equation}
\rho^a(x,x^-)\,|q^a(x,x^-)\rangle\,=\,q^a(x,x^-)\,|q^a(x,x^-)\rangle
\end{equation}
and the second is the basis of the eigenstates of the functional derivative operators
\begin{equation}
{\delta\over \rho^a(x,x^-)}\,|p^a(x,x^-)\rangle\,=\,p^a(x,x^-)\,|p^a(x,x^-)\rangle
\end{equation}
which obviously satisfy
\begin{equation}
\langle q^a|p^a\rangle\,=\,\exp\left\{i\int dx^-\,d^2x\,q^a(x,x^-)\,p^a(x,x^-)\right\}\,.
\end{equation}

The Yang vacuum does not contain gluons and is thus annihilated by the color  charge density operators. 
This obviously corresponds to the state\footnote{For the purpose of the discussion in this section the overall  normalization of the wavefunctions is not important. We will come back to this question in the next section.} $|Yang\rangle$ such that
\begin{equation}\label{yang}
\langle q|Yang\rangle\,=\,\delta[q(x,x^-)]\,; \ \ \ \ \ \ \ \ \ \ \ \ \ \ \ \langle p|Yang\rangle\,=\,1\,.
\end{equation}
This state is indeed annihilated by the action of $J_R^a(x)$ and $J_L^a(x)$, and therefore also by the action of $Q_i^a(x)$
\begin{equation}
J^a_{R(L)}(x)\,|Yang\rangle\,=\,Q_i^a(x)\,|Yang\rangle\,=\,0\,.
\end{equation}
This is however not the only state annihilated by all $Q_i^a(x)$. The other such state is an eigenstate of all $R(x)$ with the 
eigenvalue independent of $x$. We can choose this eigenvalue to be equal to one
\begin{equation}\label{rYin}
R^{ab}(x)|Yin\rangle\,=\,\delta^{ab}\,|Yin\rangle\,.
\end{equation}
Since the operator $R$ depends only on the derivatives with respect to $\rho$, the state $|Yin\rangle$ has the wave function which is independent of $\rho$
\begin{equation}\label{Yin}
\langle q|Yin\rangle\,=\,1; \ \ \ \ \ \ \ \ \ \ \ \ \ \ \ \ \ \langle p|Yin\rangle\,=\,\delta[p(x,x^-)]\,\,.
\end{equation}
The physics of $|Yin\rangle$ is precisely that of the black disk. The wavefunction does not depend on $\rho$, and thus all configurations of charge density are equally probable.
Note that the actual derivation of the KLWMIJ limit assumes that the charge density is small, thus strictly speaking $|Yin\rangle$ is outside the range of validity of the approximation. Nevertheless this state is a vacuum state of $H^{KLWMIJ}$ which appears as the remnant of the DDD symmetry of $H^{RFT}$.

Considering the JIMWLK hamiltonian we find a very similar situation. $H^{JIMWLK}$ has two vacua, which are in fact the same as the vacua of $H^{KLWMIJ}$. One state is annihilated by the dual rotation operators $\bar J^a_{R(L)}(x)$. The wave function of this state does not depend on $\alpha(x,x^-)$ and therefore by eq.(\ref{alpha}) also on $\rho^a(x,x^-)$.
This state is nothing but the black disk $|Yin\rangle$:
\begin{equation}
\bar J_{R(L)}^a(x)\,|Yin\rangle\,=\,\bar Q^a_i(x)\,|Yin\rangle\,=\,0\,.
\end{equation}
The existence of this state in the JIMWLK framework was first pointed out by Weigert in \cite{JIMWLK}.
The other vacuum state is the eigenstate of $S(x)$ with eigenvalue one. Since $S$ is the scattering matrix of a single gluon on the wavefunction in question, this state clearly corresponds to physical vacuum and up to a possible normalization constant is equal to $|Yang\rangle$.
\begin{equation}\label{syang}
S^{ab}(x)\,|Yang\rangle\,=\,\delta^{ab}\,|Yang\rangle\,.
\end{equation}
The JIMWLK Hamiltonian was derived under the assumption of large color charge density, therefore strictly speaking $|Yang\rangle$ is outside its range of validity. Nevertheless in the framework of JIMWLK evolution this state as well as states close to it (with small charge density) have been studied in the original papers \cite{JIMWLK} as well as more recently in relation to the Odderon exchange \cite{oderon}. 

\subsection{The doubling in the spectrum}
As noted above, the KLWMIJ and JIMWLK Hamiltonians separately are not self dual, thus one does not expect exact degeneracy in the spectrum of either. Nevertheless, the spectrum of each contains two towers of states built above the two vacua. To illustrate this point we will discuss the KLWMIJ evolution. The exact spectrum of the KLWMIJ Hamiltonian is not known. However in \cite{KLreggeon} we have studied it in the partonic approximation.
This approximation is obtained by expanding $R$ around unity while preserving the ultraviolet and infrared finiteness of the KLWMIJ Hamiltonian in the leading order approximation. The leading order Hamiltonian is given by
\begin{equation}
H^{part}\,=\,{\alpha_s\over 2\pi^2}\,\int d^2z \,Q^{{\rm part}\, a}_i(z)\,Q^{{\rm part}\, a}_i(z)
\label{part}
\end{equation}
with
\begin{equation}
Q^{{\rm part}\, a}_i(z)\,=\,\int d^2x{(x-z)_i\over (x-z)^2}\,{\rm tr}\left([T^a,\tilde R(x)-\tilde R(z)]\,{\delta\over\delta \tilde R^T(x)}\right)
\end{equation}
with $\tilde R=R-1$.
In this approximation we have found a "vacuum state" of the form eq.(\ref{hilbert}) with $\Sigma[R]=1$. This is just $|Yang\rangle$. We have also found eigenstates of the form
\beq\label{ef1}
G_q^\lambda\,=\,\int_x e^{iqx}\,\eta^\lambda_{cd}\,\tilde R^{cd}(x)\,|Yang\rangle
\eeq
where $\eta^\lambda_{cd}$ is a color projector onto a representation labeled by $\lambda$, and
\beq\label{G1}
G^i_q\,=\,\ P^{i\,cd}_{\,\,\,ab}\,\int_{u,v}\,\tilde R^{ab}(u)\,\tilde R^{cd}(v)\,\Psi^i_q(u,v)\,|Yang\rangle
\eeq 
with $ P^{i\,cd}_{\,\,\,ab}$ - a projector from the product of two adjoint representations onto a representation labeled by $i$. The wavefunction 
$\Psi^i_q(u,v)$ for any $i$ satisfies the BFKL equation. 

As explained in detail in \cite{KLreggeon}, 
from the $t$-channel point of view each factor of $\eta^\lambda R$ corresponds to a $t$ - channel exchange with  the color
 quantum number in representation $\lambda$. Each wavefunction in eq.(\ref{ef1}) 
describes one such exchange, whereas the wavefunctions in eq.(\ref{G1}) describe a pair of such exchanges which include 
the BFKL pomeron and odderon trajectories. 

On the other hand
when interpreted in terms of the $s$ - channel gluons, every power of $R$ in the wave function of RFT corresponds to a gluon in the QCD state.
For example the state
\beq\label{gluon}
|G^{ab}(x)\rangle\,\sim\, R^{ab}(x)|Yang\rangle 
\eeq
corresponds to the QCD state with a single gluon at the transverse position $x$ which changes its color index from $a$ to $b$ in the process of scattering. Analogously
\beq
|G^{ab,cd}(x,y)\rangle\,\sim\, R^{ab}(x)\,R^{cd}(y)|Yang\rangle 
\eeq
is the RFT representation of the QCD state with two gluons at transverse positions $x$ and $y$.
 Thus the wavefunctions of eqs.(\ref{ef1},\ref{G1}) contribute to the evolution of the scattering amplitude of a one and two gluon states respectively. These states therefore naturally belong to the sector of the Hilbert space which is close to the Yang vacuum.

Since $H^{part}$ is homogeneous in $\tilde R$ one may expect to get from it some information about the Yin part of the spectrum also.
This is indeed the case. First of all it is obvious that the Yin vacuum  is also the vacuum of $H^{part}$. More importantly it turns out that $H^{part}$ is invariant under a modified version of the DDD. In particular it is a matter of simple algebra to check that $H^{part}$ is invariant under the following transformation
\beq\label{moddual}
\tilde R(x)\,\rightarrow\, \int d^2y\,{1\over \partial^2}(x,y)\,{\delta\over \delta \tilde R(y)}; \ \ \ \ \ \ \ \ \ \ \ \ \ \ \ \ \ \ \ \ 
{\delta\over \delta \tilde R(x)}\,\rightarrow \,\partial^2\tilde R(x)
\eeq
with ${1\over \partial^2}(x,y)\,=\,{1\over 2\pi}\, \ln [(x-y)^2\mu^2]$.
This is the linearized version of the DDD transformation.

Thus we immediately infer that the states in the Yin sector of the form
\beq\label{ef2}
\tilde G_q^\lambda\,=\,\int_y \,e^{iqy}\,\eta^\lambda_{cd}\,{\delta\over \delta \tilde R^{cd}(y)}\,|Yin\rangle
\eeq
and
\beq\label{G2}
\tilde G^i_q\,=\,\ P^{i\,cd}_{\,\,\,ab}\,\,\int_{x,y,u,v}\,\ln [(u-x)^2\mu^2]\,\ln [(v-y)^2\mu^2]\,
\Psi^i_q(x,y)\,{\delta\over\delta \tilde R^{ab}(u)}\,{\delta\over\delta \tilde R^{cd}(v)}\,|Yin\rangle
\eeq   
are eigenstates of $H^{part}$.
The corresponding eigenvalues are identical to those of the corresponding states in the Yang sector.

We note that although the parton like approximation is not identical to the BFKL approximation to the KLWMIJ evolution, the invariance under the modified duality transformation is shared by the BFKL Hamiltonian.
The BFKL approximation is equivalent to expanding the factors $R(x)$ in the KLWMIJ Hamiltonian to lowest order in $\delta/\delta\rho$. 
The resulting Hamiltonian is\footnote{Though the BFKL Hamiltonian can be formally written as the sum of squares, as is well known, its
lowest eigenvalue (in our convention) is negative. The resolution of the paradox is in the fact that the BFKL eigenfunctions are
not normalizable and therefore positive definiteness of the Hamiltonian does not hold.}
\begin{equation}
H^{BFKL}\,=\,{\alpha_s\over 2\pi^2}\,\int d^2z \,Q^{{\rm BFKL}\, a}_i(z)\,Q^{{\rm BFKL}\, a}_i(z)
\end{equation}
with
\begin{equation}
Q^{{\rm BFKL}\, a}_i(z)\,=\,\int d^2x\,
{(x-z)_i\over (x-z)^2}\,f^{abc}\,\left[{\delta\over\delta \rho^b(x)}\,-\, {\delta\over\delta \rho^b(z)}\right]\,\rho^c(x)
\end{equation}
This expression is invariant under the duality transformation 
\beq\label{moddualbfkl}
{\delta\over\delta\rho^a(x)}\,\rightarrow\, \int d^2y\,{1\over \partial^2}(x,y)\,\rho^a(y); \ \ \ \ \ \ \ \ \ \ \ \ \ \ \ \ \ \ \ \ 
\rho^a(x)\,\rightarrow \,\partial^2\, {\delta\over\delta\rho^a(x)}\,.
\eeq
Thus the exact doubling of solutions corresponding to the Yin and Yang sectors exists also in the BFKL approximation.

\subsection{Holes in the black disk: the JIMWLK perspective}
The structure of two vacua and two "towers" of excitations is also supported by the JIMWLK limit. This is inevitable since the JIMWLK and KLWMIJ spectra are unitarily equivalent as the two Hamiltonians are related by the unitary DDD transformation. It is nevertheless 
instructive to analyze the spectrum from the JIMWLK viewpoint.

First, as discussed above the two vacua of JIMWLK are exactly same states as the two vacua of KLWMIJ. This time however, it is the black disk state - the $|Yin\rangle$ which is the natural vacuum of the JIMWLK evolution, while the presence of $|Yang\rangle$  is "accidental" and is the remnant of the DDD symmetry of $H^{RFT}$.

To get an idea about the nature of the eigenstates we can employ the approximation dual to the parton model eq.(\ref{part}). 
\begin{equation}
\tilde H^{part}\,=\,{\alpha_s\over 2\pi^2}\,\int d^2z\, \tilde Q^{{\rm part}\, a}_i(z)\,\tilde Q^{{\rm part}\, a}_i(z)
\label{part1}
\end{equation}
with
\begin{equation}
\tilde Q^{{\rm part}\, a}_i(z)\,=\,\int d^2x{(x-z)_i\over (x-z)^2}\,{\rm tr}\left([T^a,\tilde S(x)-\tilde S(z)]\,{\delta\over\delta \tilde S^\dagger(x)}\right)
\end{equation}
with $\tilde S=S-1$.
The eigenstates of this Hamiltonian are obviously DDD duals of the "reggeon" and "pomeron" states found in \cite{KLreggeon}.
The states of the Yin "tower" are
\begin{eqnarray}\label{efh1}
H_q^\lambda\,&=&\,\int_x \,e^{iqx}\,\eta^\lambda_{cd}\,\tilde S^{cd}(x)\,|Yin\rangle\\
H^i_q\,&=&\,\ P^{i\,cd}_{\,\,\,ab}\,\int_{u,v}\,\tilde S^{ab}(u)\,\tilde S^{cd}(v)\,\Psi^i_q(u,v)\,|Yin\rangle\nonumber
\end{eqnarray} 
and the like, while the states in the Yang tower are
\begin{eqnarray}\label{efh2}
\tilde H_q^\lambda\,&=&\,\int_y \,e^{iqy}\,\eta^\lambda_{cd}\,{\delta\over \delta \tilde S^{cd}(y)}\,|Yang\rangle\\
\tilde H^i_q\,&=&\,\ P^{i\,cd}_{\,\,\,ab}\,\,\int_{x,y,u,v}\ln [(u-x)^2\mu^2]\ln [(v-y)^2\mu^2]\,\Psi^i_q(x,y)
\,{\delta\over\delta \tilde S^{ab}(u)}\,{\delta\over\delta \tilde S^{cd}(v)}\,|Yang\rangle
\end{eqnarray}
etc.

The interpretation of the states of eq.(\ref{efh1}) is quite amusing. Consider for example the dual of a single gluon state eq.(\ref{gluon}). 
\beq\label{hole}
|H^{ab}(x)\rangle\,=\,S^{ab}(x)\,|Yin\rangle \,.
\eeq
This state is black almost everywhere except at the transverse position $x$.
To see this imagine that we scatter on this state ($W\,=\,|H^{ab}(x)\rangle$) 
a single gluon projectile at transverse position $z$. The projectile averaged single gluon $S$-matrix is simply $\Sigma[S]=S^{cd}(z)$. To calculate the scattering matrix for such a process we have to integrate this $\Sigma$ with the weight function $W$ corresponding to the state eq.(\ref{hole})
\footnote{Eq.(\ref{hole}) in fact defines an $SU(N)\otimes SU(N)$ multiplet of states corresponding to the indices $ab$. The same is true for the projectile averaged $S$-matrix . For scattering of a gluon with a given color on a given state we should specify all four indices appropriately. We will however keep the indices arbitrary as it does not cause any additional complications.}
\beq
{\cal S}\,=\,\int DS\, S^{cd}(z)\,S^{ab}(x)\,=\,{1\over (N^2-1)}\,\delta_{x-z}\,\delta^{ac}\,\delta^{bd}
\eeq
with the "Kronecker delta": $\delta_{0}=1$. 
Thus multiplication of the black disk state $| Yin\rangle$ by a factor $S(x)$ produces the state which is not completely black for a gluon that impinges on it at the point $x$. 
\begin{figure}[htbp]
\centerline{\epsfig{file=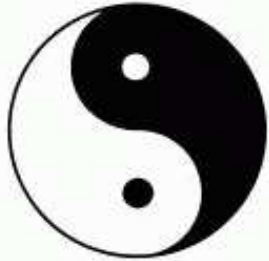,width=40mm}}
\caption{\it The Yin-Yang structure: duality between gluon excitations above the white vacuum and the hole excitations above the black vacuum. }
\label{fig1}
\end{figure}
It is natural to think of such a state as a hole in the black, see Fig. 1. We will adopt this terminology and will call $S(x)$ - a hole creation operator. Thus the excitations in the $|Yin\rangle$ sector can be though of as the "reggeized holes" and their bound states - the "hole Pomerons".
Even though formally from the point of view of the eigenstates of $H^{RFT}$ there is complete symmetry between the gluon and the hole states, physically of course they are very different and they contribute quite differently to physical amplitudes.
 We now turn to the discussion of this question.

\section{The $RFT$ Hilbert space and the physical states}

To calculate physical amplitudes using eq.(\ref{over}) we need to understand how to expand a weight functional $W$ that describes a physical state in the basis of eigenfunctions of $H^{RFT}$. As mentioned above the functional $W$ cannot be arbitrary but rather must satisfy some conditions. This issue has been discussed before in \cite{KLreggeon} and \cite{kl5}. We repeat the main points here since they are important for our subsequent discussion.

\subsection{Normalization of physical states}
First, as shown in \cite{kl5} in order for the correlators of powers of $\rho^a$ to be consistent with the quantum commutation relations of the operators $\hat\rho^a$ the functionals $W$ must be of the form
\begin{equation}\label{hilbert}
W[\rho]\,=\,\Sigma[R]\,\delta[\rho(x,x^-)]
\end{equation}
with an arbitrary functional $\Sigma$ which depends  only on the charge shift operator or the "dual Wilson line" $R$.
Eq.(\ref{hilbert}) is a restriction on the Fourier transform of $W$. Inverting it we find
\begin{equation}\label{hilbert1}
\int D\rho\, \exp\left\{i\int dx^-d^2x\,\rho^a(x,x^-)\,\alpha^a(x,x^-)\right\}\,\,W[\rho]\,=\,\Sigma[S]
\end{equation}
with $S$ defined as in eq.(\ref{S}). Thus the Fourier transform of $W$ must depend only on the Wilson line $S$ and not on any other combination of the conjugate variables $\alpha$.  This has a transparent physical interpretation. The functional Fourier transform of $W$ is precisely the projectile averaged scattering matrix for scattering on the "external field" $\alpha$. The restriction eq.(\ref{hilbert}) simply means that this scattering matrix must be a function of scattering matrices of individual gluons constituting the projectile, and does not depend on any other property of the external field. The normalization of the functional $W$ is determined by requiring that for $\alpha=0$ the scattering matrix is equal to unity. This leads to the overall normalization condition
\begin{equation}\label{norm}
\Sigma[R=1]\,=\,1,\ \ \ \ \ \ \ \ \ \ \ \ \ \ \
 \int D\rho\, W[\rho]\,=1\,.
\end{equation}

Further general properties of $\Sigma$ follow from its identification as the projectile averaged scattering matrix. Suppose for example the wave function of the projectile at the transverse position $x$ has the form (we take there to be exactly one gluon at the point $x$)
\begin{equation}\label{psi1}
|\Psi_x\rangle\,=\,\sum_{a=1}^{N^2-1}\,C_a(x)\,|a,x\rangle\,.
\end{equation}
The scattering matrix operator when acting on $\Psi$ multiplies the gluon wave function by the matrix $S$ \cite{kl1}. 
We thus have
\begin{equation}
\Sigma[S]\,=\,\langle\Psi_x|\hat S|\Psi_x\rangle\,=\,\sum_{a,b=1}^{N^2-1}\,C_a(x)\,C^*_b(x)\,\,S^{ab}(x)\,.
\end{equation}
This generalizes to states with more than one gluon. Expanding the functional $\Sigma$ in Taylor series in $R$ we obtain the general form compatible with its interpretation as the projectile averaged scattering matrix
\begin{equation}\label{sigmas}
\Sigma[R]\,=\,\sum_{m=0}^{\infty}\sum_{n_i=1}^{\infty}\left[C_{\{a^1\}...\{a^m\}}(x_1,...,x_m)C^*_{\{b^1\}...\{b^m\}}(x_1,...,x_m)\right]
\Pi_{i=1}^{m}\left[R^{a^i_1b^i_1}(x_i)...R^{a^i_{n_i}b^i_{n_i}}(x_i)\right]
\end{equation}
where $\{a^1\}$ denotes the set $a^1_1,a^1_2,...,a^1_{n_1}$, etc.
Simply stated this is the $S$-matrix of a state whose quantum mechanical amplitude to have $n_i$ gluons with color indices
$a^i_1,...,a^i_{n_i}$ at the transverse coordinate $x_i$ with $i=1,...,m$ is $C_{\{a^1\}...\{a^m\}}(x_1,...,x_m)$.
Thus for example the coefficients of "diagonal" terms with $a^i_j=b^i_j$ must be positive, since they have the meaning of probabilities to find the configuration with particular color indices in the projectile wave function.
One should note that in the derivation of eq.(\ref{sigmas}) we have assumed that the projectile is in a pure state and is described by a wave function. If instead the projectile is described by a mixed density matrix a wider set of functions $\Sigma$ is allowed. We will not enter into a more detailed discussion here as we will not have to use these conditions below. 

As is obvious from the discussion above, the normalization conditions on physical states are 
very different than the conventional normalization of states in the Hilbert space of RFT eq.(\ref{hnorm}).

Next consider the eigenvalue equation of the JIMWLK Hamiltonian
\beq\label{eig}
H^{JIMWLK}\,\Psi_i\,=\,\omega_i\,\Psi_i\,.
\eeq
Since $H^{JIMWLK}$ is proportional to the generator of rotations, integrating this over the group with any group invariant measure leads to
\beq\label{t1}
\int D\rho\,\Psi_i[S]\,=\,0, \ \ \ \ \ \ \ \ {\rm for}\ \ \ \ \ \ \omega_i\,\ne\,0
\eeq
Another property of eq.(\ref{eig}) is that its left hand side vanishes at $S=1$.
Thus we also have 
\beq\label{t2}
\Psi_i[S=1]\,=\,0, \ \ \ \ \ \ \ \ {\rm for}\ \ \ \ \ \ \ \omega_i\,\ne\,0\,.
\eeq
Since the eigenfunctions of $H^{KLWMIJ}$ are functional Fourier transforms of the eigenfunctions of $H^{JIMWLK}$, and since eqs.(\ref{t1}) and (\ref{t2}) are Fourier transforms of each other, we conclude that the eigenfunctions of $H^{KLWMIJ}$ share these properties. This can of course be seen by considering directly the eigenvalue equation for the KLWMIJ eigenfunctions. 

Eqs.(\ref{t1},\ref{t2}) also hold for the wavefunctions of the complete $H^{RFT}$. Eq.(\ref{t1}) is simply the statement that all eigenstates with nonvanishing eigenvalues are orthogonal to the Yin vacuum, while eq.(\ref{t2}) is the orthogonality condition of these states with the Yang vacuum. The property of orthogonality of eigenstates corresponding to  different eigenvalues has to hold for eigenfunctions of any hermitian operator, and thus also $H^{RFT}$.

Eq.(\ref{t1}) in conjunction with the normalization of the physical wave functions eq.(\ref{norm}) has an immediate consequence that any physical state must have a nonvanishing projection onto a state with zero energy.
We know that $H^{RFT}$ has two zero energy states - Yin and Yang. We will assume for the purpose of the current discussion that those are the only two states with zero energy. We will discuss the possibility of existence of additional zero energy states in the next subsection. 

We now have to discuss in more detail the normalization of the Yin and Yang eigenfunctions. To write the properly normalized wave functions in a sensible way we have to introduce the infrared and ultraviolet cutoff on the system - the total volume $V$ and the closest distance $1/\Lambda$ in the transverse plane. 
We also have to introduce the "infrared" and "ultraviolet" cutoffs in the field space. In particular we will cutoff the $\rho$ integrals at some large value $v$ and at a small value $\Delta$. With these cutoffs in place we can revise eq.(\ref{yang},\ref{Yin}) appropriately
\begin{eqnarray}\label{normalized}
G_0\,=\,\langle q|Yang\rangle&=&\Pi_x \,\Delta^{1/2}\,\delta(q(x)); \\
H_0\,=\,\langle q|Yin\rangle&=&\Pi_x \,v^{-1/2}\,.
\end{eqnarray}
Since we would like to keep the DDD explicit at finite cutoff, the two cutoffs $\Delta$ and $v$ are not independent. In particular, we want the state $H_0$ to be the Fourier transform of the state $G_0$. This requires $\Delta=v^{-1}$.
Thus for the normalized eigenstates we have
\beq
\int D\rho\,G_0^2\,=\,\int D\rho\,H_0^2\,=\,1\,;\ \ \ \ \ \ \ \ \ \ 
\int D\rho\,G_0\,=\,v^{-{n/ 2}}; \ \ \ \ \ \ \ \ \ \ \ 
\int D\rho\,H_0\,=\,v^{{n/ 2}}
\eeq
where $n=V\Lambda$ is the total number of points in the transverse plane. 

With this normalization the expansion of a physical state in the eigenfunctions of $H^{RFT}$ has the form\footnote{Compared to \eq{basw}
we explicitly take into account the doubling of the spectra. Namely the index $l_i$ runs up to two: gluons and holes.}
\beq
W[S]\,=\,\gamma_0\,v^{{n/ 2}}\,G_0\,+\,\eta_0\,v^{-{n/ 2}}\,H_0\,+\,
v^{{n/ 2}}\,\sum_{i:\,\omega_i> 0}\,\gamma_i\, G_i[S]\,+\,v^{-{n/ 2}}\,\sum_{i:\,\omega_i> 0}\,\eta_i\,H_i[S]
\eeq
with
\beq\label{total}
\gamma_0\,+\,\eta_0\,=\,1
\eeq
which follows from the overall normalization (\ref{norm}). The states $G_i$ and $H_i$ are "gluon" and "hole" exited states belonging to the Yang and Yin
Hilbert spaces respectively.
Assuming that the normalization of other states in the Yang tower $G_i$ is similar to that of $G_0$ as far as the volume and ultraviolet cutoff is concerned and the same for $H_i$ relative to $H_0$, we expect all coefficients $\gamma_i$ and $\eta_i$ to be numbers of order unity.
For $\Sigma$, which is the Fourier transform of $W$ the expansion is analogously
\beq
\Sigma[S]\,=\,\gamma_0\,v^{{n/ 2}}\,H_0\,+\,\eta_0\,v^{-{n/ 2}}\,G_0\,+\,
v^{{n/ 2}}\,\sum_{i:\,\omega_i> 0}\,\gamma_i\, H_i[S]\,+\,v^{-{n/2}}\,\sum_{i:\,\omega_i> 0}\,\eta_i\,G_i[S]\,.
\eeq
Note that although the Yin and Yang sectors form separate Hilbert spaces in the thermodynamic limit,
 at finite value of the regulators the overlap between the states in the two sectors is finite. In particular
\beq
\langle Yin|Yang\rangle\,=\,
\int D\rho \,G_0\,H_0\,=\,v^{-n}\,.
\eeq

Thus for scattering of a projectile $P$ on a target $T$ we have
\begin{eqnarray}
{\cal S}(Y)&=&(\gamma^P_0\,+\,\eta^P_0)\,(\gamma^{T}_0\,+\,\eta^{T}_0)\,-\,{v^n\,-\,v^{-n}\over v^n}\,\eta^P_0\,\eta^{T}_0+\\
&+&\sum_{i:\,\omega_i> 0}e^{-\,\omega_i\, Y}\,\left[\gamma^P_i\,\eta^{T}_i\,+\,\eta^P_i\,\gamma^{T}_i\,+\,
\left(v^n\,\gamma^P_i\,\gamma^{T}_i\,+\,v^{-n}\,\eta^T_i\,\eta^P_i\right)\,\int D\rho \,G_i\,H_i\right]\nonumber
\end{eqnarray}
or in the infinite volume limit
\beq\label{infi}
{\cal S}(Y)\,=\,1\,-\,
\eta^P_0\,\eta^{T}_0\,+\,
\sum_{i:\,\omega_i> 0}\,e^{-\,\omega_i\, Y}\,
\left[\gamma^P_i\,\eta^{T}_i\,+\,\eta^P_i\,\gamma^{T}_i\,+\,\gamma^P_i\,\gamma^{T}_i\,v^n\,\int D\rho\, G_i\,H_i\right]\,.\nonumber
\eeq
 
This formula tells us that if neither the projectile nor the target have an overlap with the white vacuum 
($\gamma_0^T=\gamma_0^P=0$; $\eta_0^T=\eta_0^P=1$), the $S$ - matrix at large rapidity vanishes, as only the positive energy eigenstates 
contribute.
If either the projectile or the target have a white component, this means that $\eta_0^P\eta_0^T\ne 1$, and therefore the $S$ matrix tends to a constant value as $Y\rightarrow\infty$. This is the situation when the projectile (or the target) has a vacuum component in its wave function - this component remains transparent even at asymptotically high rapidity. This of course is not the situation one usually is interested in.
It is in fact easy to see that physical states that contain finite number of gluons are all orthogonal to the $|Yang\rangle$ state.
For example a single gluon state (eq.(\ref{gluon})) properly normalized to satisfy eq.(\ref{norm})
\beq
\langle Yang|G^{ab}(x)\rangle\,=\,v^{n/2}\,\langle Yang|R^{ab}(x)|Yang\rangle \,=\,0
\eeq
by virtue of $|Yang\rangle$ being invariant under the $SU(N)\times SU(N)$ group that rotates the left and right indices of $R$. The same argument applies to any state which contains a nonzero number of gluons at some transverse position. On the other hand all these states have the same overlap with the $|Yin\rangle$ vacuum
\begin{eqnarray}
\langle Yin|G^{ab}(x_1)...G^{cd}(x_n)\rangle&=&v^{n/2}\,\langle Yin|R^{ab}(x_1)...R^{cd}(x_n)|Yang\rangle\,=\\
&=&v^{n/2}\,\delta^{ab}...\delta^{cd}\,\langle Yin|Yang\rangle\,=\,v^{-{n/ 2}}\,\delta^{ab}...\delta^{cd}\,.\nonumber
\end{eqnarray}
Thus for all physical states of interest $\gamma_0=0$ and $\eta_0=1$.

\subsection{Tertium non data}
The existence of the white and black vacua of $H^{RFT}$ is not in doubt. An interesting question is whether the theory could have an additional "vacuum" $\omega=0$ state. Such a state would be selfdual and presumably "grey" in the sense that it would lead to $S$ matrix not equal to zero or unity. 

At first sight the existence of such a state is suggested by the gluon-hole duality itself. 
Indeed, we know that the state which contains initially a small number of gluons becomes darker as the result of the evolution. This darkening is illustrated for example by the growth of the number of gluons. The gluon number operator in the framework of the BFKL evolution is simply related to the Fourier transform of the charge density operator
\beq
N_g(k)\propto {1\over k^2}\langle \rho^a(k)\rho^a(-k)\rangle\,.
\eeq
Within KLWMIJ evolution the charge density correlator satisfies the BFKL equation and thus grows exponentially with rapidity. The number of gluons therefore grows exponentially fast and the state becomes blacker. 

What happens if instead one starts with an initial state close to the black disk limit and follows the evolution of the number of holes? The observable corresponding to the number of holes clearly is just the dual of the number of gluons
\beq\label{holen}
N_h(k)\propto {1\over k^2}\langle {\delta\over\delta\alpha^a(k)}{\delta\over\delta\alpha^a(-k)}\rangle\,.
\eeq
The rapidity evolution close to the black disk limit is given by the JIMWLK equation. This leads to a linear BFKL evolution of the correlator in eq.(\ref{holen}) in a manner totally analogous to the linear evolution of the $\rho$ correlators in KLWMIJ.
So we conclude that close to black disk the number of holes exponentially grows with the BFKL exponent. 

It is tempting to conclude from this argument that the black disk is an unstable state and if one chooses an initial condition sufficiently close to it, this initial state will flow away from the black disk in the course of the evolution. The final destination for such a state 
then may be some state other than either black or white vacuum - a "grey vacuum".
If such a state exists, one would expect its properties to be quite interesting. First, it would be invariant under the DDD transformation. The gluon operator is equal to unity when acting on the black disk state, eq.(\ref{rYin}). In this sense the black disk can be thought of as a condensate of gluons, which breaks the $SU(N)\otimes SU(N)$ gluon rotation symmetry down to the diagonal $SU_V(N)$. The white state on the other hand according to eq.(\ref{syang}) is the condensate of holes and breaks the hole rotation symmetry in the same fashion. One could expect the grey state to be selfdual, break no symmetries and have neither gluon nor hole condensates. In physical terms that would mean that the scattering probability of say a dipole of size $r$ will be equal neither to zero nor to unity for any finite size $r$. One would of course still expect the probability to approach unity for $r\rightarrow\infty$ and zero for $r\rightarrow 0$, but only asymptotically. The most natural realization of such properties would be if the state was conformally invariant and yielded correlation functions which behaved as a power of the distance. We note that these properties are consistent with the asymptotic behavior suggested in \cite{levin}. Such a selfdual state also exists on the classical level in Gribov's Reggeon field theory \cite{Gribov, Amati, BF}, although its fate beyond classical approximation is unclear\footnote{The conclusion of \cite{BF} e.g. is that the asymptotic behavior is indeed grey but not related to the classical selfdual solution $p=q=0$.}.

The hole proliferation argument however cannot be interpreted directly as pointing to the existence of an additional vacuum. The hole states crucially differ from gluons in that they do not satisfy the conditions of physicality. In particular any state with finite number of holes has a zero norm in the sense of eq.(\ref{norm}), since it is by definition orthogonal to the $|Yin\rangle$ vacuum. Since the condition of eq.(\ref{norm}) expresses the normalization of the QCD wave function of the projectile (or target), zero norm states necessarily contain negative probabilities and thus are clearly unphysical. A complementary feature of the hole states is that they are not orthogonal to the $|Yang\rangle$ state. Thus even if physical interpretation were possible, the QCD state involved would contain a vacuum component in its wave function. It would not be therefore of the type of state which one expects to get black as the result of high energy evolution. The proliferation of holes may be simply a manifestation of the fact that the "white" component in the hole wavefunction dominates the $S$ matrix at high energies.

Indeed, let us present an argument to the effect that the grey state if it exists does not play any role in the evolution of physical states. We have shown in the previous subsection that any physical state, namely any state consisting of a finite number of gluons has unit projection on the $|Yin\rangle$ state, $\eta_0=1$. It therefore follows from eq.(\ref{total}) that its projection onto any other state with unit physical norm defined by eq.(\ref{norm}) vanishes\footnote{Eq.(\ref{total}) assumes existence of two vacua only, but its generalization to a greater number of such states is straightforward.}. As regards the grey vacuum this means that either a) it does not exist, or b) it exists but its overlap with any physical state vanishes, or c) its physical norm vanishes, 
in which case it cannot be a final point for evolution of any physical state. Our present argument does not exclude the possibility c), namely that the finite point of the evolution is an admixture of the black vacuum and a zero physical norm "grey vacuum". We view this situation as highly improbable, since in this case the "grey vacuum" itself is not a physical state, unlike the Yin and Yang states. We believe this possibility can be excluded by consideration of other restrictions on the nature of physical states as encoded in eq.(\ref{sigmas}).
With this proviso we conclude that a grey state even if it exists in the Hilbert space of $H^{RFT}$ does not couple to evolution of physical states\footnote{This argument seemingly contradicts conclusions of \cite{levin} as well as those of the recent paper \cite{genyakozlov1}, which finds a grey state as the final state of the high energy evolution in a toy model without transverse dimensions. We note however that the properties of the evolution in the toy model of \cite{genyakozlov1} differ in a crucial way from those in QCD. In particular the Hamiltonian of \cite{genyakozlov1} is not hermitian, thus many of our arguments do not apply to this case. Also, and perhaps more importantly, as explained in \cite{genyakozlov1} the toy model Hamiltonian describes real processes of both emission and annihilation of dipoles in the wave function. In such a situation one indeed expects that the evolution should terminate on a grey state, since the dipole density in the wave function must literally saturate. In QCD on the other hand only emission of gluons is present but annihilation is not allowed, see \cite{kl5} for discussion. Thus one does not expect saturation of the gluon density in the QCD wave function, although the rate of its growth is expected to slow down dramatically at high energy.}.

\section{Discussion}
In this paper we have discussed the structure of the Hilbert space of $H^{RFT}$, the realization of the DDD symmetry and some general consequences of this for the high energy evolution of physical states. 
We have shown that the DDD symmetry is spontaneously broken leading to appearance of two zero "energy" states - the white "Yang" vacuum and the black disk "Yin" vacuum. The excitations in the Yang Hilbert space have the character of gluons (or rather reggeized gluons) and multigluon states. The dual of these states in the Yin Hilbert space can be thought of as holes in the black disk (or reggeized holes to complete the analogy). The gluon "creation operator" in the framework of the RFT is the color charge density shift operator $R$. It has the double meaning of creating the charge density of a gluon in the hadronic wave function and also as the scattering matrix of the gluon belonging to the projectile when it scatters on an external hadronic target. The hole operator $S$ is the scattering matrix of a gluon impinging on the hadronic wave function. 

The gluon and hole operators are dual with respect to DDD and fit the standard pattern of order-disorder variables taken to the extreme. The operator $R(x)$ creates an extra adjoint charge at the point $x$, while the operator $S(y)$ measures the color field created by this charge. Perturbatively this field is $\alpha(y)\propto g\ln(y-x)^2\mu^2$ where $\mu$ is the infrared cutoff. Thus the phase of operator $S(y)$ created by the action of $R(x)$ is of order $\alpha_s\ln(y-x)^2\mu^2$. At distances $|y-x|<r_c=1/\mu e^{1/\alpha_s}$ this phase is very large. The operator $R$ therefore disorders any configuration of $S$ on the distance scale $r_c$. Since the scale $r_c$ itself is proportional to the infrared cutoff, this is an extremely strong disordering effect. 

The Yin and Yang vacua fit naturally in the picture of order-disorder duality. The hole operator $S$ is equal to unity when acting on the Yang vacuum, thus holes are condensed in $|Yang\rangle$. Likewise, $|Yin\rangle$ is the condensate of gluons. The symmetry breaking pattern in the two states is also different. The gluon operator transforms as an adjoint representation under the $SU_g(N)\otimes SU_g(N)$ transformation as $R\rightarrow U_g\,R\,\tilde U_g$. Although we do not know the full $H^{RFT}$ Hamiltonian. we do know that this transformation is a symmetry of its KLWMIJ limit. Analogously the JIMWLK Hamiltonian is invariant under the $SU_h(N)\otimes SU_h(N)$ transformation $S\rightarrow U_h\,S\,\tilde U_h$. The vector subgroup of these two groups is the same, since under it both $\rho^a$ and $\delta/\delta\rho^a$ transform as adjoint representation. However other group elements are different. In particular in the Yin vacuum the gluon symmetry $SU_g(N)\otimes SU_g(N)$ is broken down to the diagonal $SU_V(N)$ but the hole symmetry $SU_h(N)\otimes SU_h(N)$ remains unbroken. In the Yang vacuum the situation is reversed. This situation is generic for theories involving order-disorder variables. Some symmetries which are represented {\it a la} Noether for one set of variables look topological when written in terms of the dual set, and vice versa. Although we have not found an explicit realization of say $SU_h(N)\otimes SU_h(N)$ in terms of gluons, we may expect that the elements which do not belong to the vector subgroup have topological interpretation.

The spontaneous breaking of these global symmetries should lead to the appearance of the Goldstone boson modes. Indeed in the framework of the parton like approximation to the KLWMIJ equation these Goldstone modes were identified with the reggeized gluons. By duality the reggeized holes are the Goldstone modes in the JIMWLK case. Whether this situation persists in the full RFT is at this point unclear. The full $H^{RFT}$ is selfdual. Thus either both $SU(N)\otimes SU(N)$ groups are exact symmetries of the full RFT and the spontaneous breaking pattern in the vacua persists, or only the common diagonal subgroup remains the symmetry of $H^{RFT}$. In the latter case the Goldstone bosons only appear in the dense and dilute limits and are artifacts of these limiting cases.

Although our discussion has been rather formal, we have been able to draw some interesting conclusions. In particular we were able to argue that the high energy asymptotics in the full $H^{RFT}$ must be black and not grey.
The formula eq.(\ref{infi}) is also informative. For physical states ($\eta=1$) the $S$ matrix is given by
\beq\label{infi1}
{\cal S}(Y)\,=\,\sum_{i:\,\omega_i> 0}\,e^{-\,\omega_i \,Y}\,
\left[\gamma^P_i\,\eta^{T}_i\,+\,\eta^P_i\gamma^{T}_i\,+\,\gamma^P_i\,\gamma^{T}_i\,v^n\,\int D\rho\, G_i\,H_i\right]\,.\nonumber
\eeq
One can infer from it for example that the approach of the $S$ matrix to the black disk limit must be exponential. At large $Y$ the leading contribution comes from the lowest eigenvalue that couples to the particular physical state. In principle in the case of a gapless spectrum this eigenvalue could be zero leading to a power like  approach to the black disk limit. However if the only massless modes are the Goldstone bosons of the type discussed above, and therefore carry color quantum number, they will decouple from the $S$ - matrix of any color neutral physical state. Their presence could nevertheless be felt in that the spectrum even at finite $\omega$ can be continuous. From this perspective we view the recent result about the asymptotics of the solutions of the Kovchegov equation \cite{genyakozlov} as very natural. It was shown in \cite{genyakozlov} that the approach of the $S$ 
matrix of a dipole to the black disk limit does not follow a gaussian ${\cal S}
\propto e^{-\epsilon Y^2}$ as suggested in \cite{levintuchin}, but is rather exponential with a power prefactor,
${\cal S}\propto Y e^{-\omega_0Y}$. From the perspective of eq.(\ref{infi1}) this corresponds to the continuous 
spectrum with threshold at 
$\omega_0$ and density of states of the form $1/(\omega-\omega_0)$.

Eq.(\ref{infi1}) is also interesting from the following point of view. We expect that if the hadron at initial rapidity contains a small number of gluons, its wavefunction will have large projection mostly on the states in the Yang space, $\eta_i\approx 0$. Analogously if the hadron is very dense, it has a large projection onto the Yin space, $\gamma_i\approx 0$. Thus if we consider scattering of a dilute projectile on a dense target or vice versa, only one of the first two terms in the sum in eq.(\ref{infi1}) contributes, while the third term is irrelevant. This corresponds to the situation described by the KLWMIJ or JIMWLK evolution. In this case one does not need to know the off diagonal matrix elements between the Yin and Yang eigenstates. The third term is important only if we are in the situation when both hadrons involved in the scattering are initially dilute.  This is of course precisely the situation in which we expect the Pomeron loops to give important contribution to the evolution at high enough rapidity. Thus the last term in eq.(\ref{infi1}) is naturally associated with the contribution of Pomeron loops. 

This observation suggests an approximate way to include the Pomeron loops in the calculation of the scattering matrix via eq.(\ref{infi1}). Namely rather than searching for the full hamiltonian $H^{RFT}$ one could solve for the eigenfunctions of $H^{JIMWLK}$ in the Yin space, $H_i$ and for the eigenfunctions of $H^{KLWMIJ}$ in the Yang space, $G_i$. One can then use these two sets to approximate the third term in the sum of eq.(\ref{infi1}). In spirit this is similar to approximations discussed in the recent literature \cite{IT,MSW,LL3,kl,levin} which strive to keep both the "Pomeron splittings" and "Pomeron mergings" in the evolution of the scattering amplitude. The advantage of the present  suggestion is that it treats both processes symmetrically, and also keeps the essential nonlinearities in calculating each set of the eigenfunctions. It does not take into account any perturbative corrections to the eigenfunctions themselves, which will certainly be present in $H^{RFT}$. One may hope however that these corrections to the eigenfunctions $H_i$ and $G_i$ are indeed perturbatively small and negligible as long as the colliding particles at initial rapidity are perturbatively dilute. To pursue this route one would have to learn how to find the eigenfunctions and eigenvalues of the JIMWLK-KLWMIJ evolution, itself a formidable task. The potential payoff however seems to be worth
investing some effort in this direction.


\begin{thebibliography}{99}

\bibitem{Gribov}
  V.~N.~Gribov,
  Sov.\ Phys.\ JETP {\bf 26}, 414 (1968)
  [Zh.\ Eksp.\ Teor.\ Fiz.\  {\bf 53}, 654 (1967)].


\bibitem{Amati}
  D.~Amati, M.~Le Bellac, G.~Marchesini and M.~Ciafaloni,
  Nucl.\ Phys.\ B {\bf 112}, 107 (1976).
\\
V.~Alessandrini, D.~Amati and M.~Ciafaloni,
  Nucl.\ Phys.\ B {\bf 130}, 429 (1977).
\\
  D.~Amati, G.~Marchesini, M.~Ciafaloni and G.~Parisi,
  Nucl.\ Phys.\ B {\bf 114}, 483 (1976).
\\
  M.~Ciafaloni and G.~Marchesini,
  Nucl.\ Phys.\ B {\bf 109}, 261 (1976).
  
  
\bibitem{Baker}
  M.~Baker and K.~A.~Ter-Martirosian,
  Phys.\ Rept.\  {\bf 28}, 1 (1976).


\bibitem{GLR}  L.V.~Gribov, E.~Levin and M.~Ryskin, Phys. Rep.
100:1,1983.

\bibitem{MUQI}
A. H. Mueller and J. Qiu,  Nucl. Phys. {\bf B 268} (1986) 427.


\bibitem{Mueller} A. Mueller, {\it 
Nucl. Phys.} {\bf B335} 115 (1990); {\it ibid} {\bf B 415}; 
373 (1994); {\it ibid} {\bf B437} 107 (1995).


\bibitem{mv} L.~McLerran and R.~Venugopalan, Phys. Rev. D49:2233-2241,1994 
; Phys.Rev.D49:3352-3355,1994. 




\bibitem{balitsky} I. Balitsky, {\it Nucl. Phys.}  {\bf B463} 99 (1996); 
{\it Phys. Rev. Lett.} {\bf 81} 2024 (1998); 
{\it Phys. Rev.}{\bf D60} 014020 (1999).


\bibitem{Kovchegov}
  Y.~V.~Kovchegov,
  Phys.\ Rev.\ D {\bf 61}, 074018 (2000)
  [arXiv:hep-ph/9905214].

\bibitem{JIMWLK} J. Jalilian Marian, A. Kovner, A.Leonidov and H.
Weigert,
{\it Nucl. Phys.}{\bf  B504} 415 (1997); 
{\it Phys. Rev.} {\bf D59} 014014 (1999); 
J. Jalilian Marian, A. Kovner and H. Weigert, {\it Phys. Rev.}{\bf D59} 
014015 (1999); 
A. Kovner and J.G. Milhano, {\it Phys. Rev.} {\bf D61} 014012 (2000) .
 A. Kovner, J.G. Milhano and H. Weigert,
{\it Phys.Rev.} {\bf D62} 114005 (2000); 
 H. Weigert, {\it Nucl.Phys.} {\bf A 703} (2002) 823.
 

\bibitem{cgc}  E.Iancu, A. Leonidov and L. McLerran, {\it Nucl. Phys.} 
{\bf A 692} (2001) 583; {Phys. Lett.} {\bf B
510} (2001) 133;
E. Ferreiro, E. Iancu, A. Leonidov, L. McLerran;  
{\it Nucl. Phys.}{\bf A703} (2002) 489.

\bibitem{kl}  A. Kovner and M. Lublinsky, Phys.\ Rev.\ D {\bf 71}, 085004 (2005).

\bibitem{KLreggeon} A. Kovner and M. Lublinsky,  hep-ph/0512316.



\bibitem{BF}
  B.~Blok and L.~Frankfurt,
  e-print archive:hep-ph/0508218;  Mod.Phys.Lett.A21:549-558,2006, e-print archive: hep-ph/0512225.

\bibitem{BFKL}
 E. A. Kuraev, L. N. Lipatov, and F. S. Fadin,  Sov. Phys. JETP
                {\bf 45} (1977) 199 ; \\
Ya. Ya. Balitsky and L. N. Lipatov,
               {  Sov. J. Nucl. Phys.}\, {\bf 28} (1978) 22.

\bibitem{LipatovRgluon}
  L.~N.~Lipatov,
  Sov.\ J.\ Nucl.\ Phys.\  {\bf 23}, 338 (1976)
  [Yad.\ Fiz.\  {\bf 23}, 642 (1976)].

\bibitem{something} A. Kovner and M. Lublinsky, Phys.\ Rev.\ Lett.\  {\bf 94}, 181603 (2005).


\bibitem{balitsky1} I. Balitsky, in *Shifman, M. (ed.): At the frontier of particle physics, vol. 2* 1237-1342; 
e-print arxive hep-ph/0101042.  


\bibitem{shoshi1} A. Mueller and A.I. Shoshi, Nucl. Phys. B {\bf 692}, 175 (2004).

\bibitem{IMM} E. Iancu, A.H. Mueller and S. Munier, Phys.\ Lett.\ B {\bf 606}, 342 (2005).

\bibitem{IT}  E. Iancu and  D. N. Triantafyllopoulos,  Phys.\ Lett.\ B {\bf 610}, 253 (2005); 
Nucl.\ Phys.\ A {\bf 756}, 419 (2005); E.~Iancu, G.~Soyez and D.~N.~Triantafyllopoulos,
  arXiv:hep-ph/0510094.


\bibitem{kl4} A.~Kovner and M.~Lublinsky,
Phys.\ Rev.\ D {\bf 72}, 074023 (2005)
  [arXiv:hep-ph/0503155].



\bibitem{MSW}A. Mueller, A. Shoshi and S. Wong, Nucl.\ Phys.\ B {\bf 715}, 440 (2005).

\bibitem{LL3} E. Levin and M. Lublinsky; Nucl.Phys.A763:172-196,2005 , e-print arxive hep-ph/0501173.

\bibitem{levin} E. Levin,  Nucl.Phys.A763:140-171,2005; 
e-Print Archive: hep-ph/0502243.

\bibitem{kl1} A. Kovner and M. Lublinsky, JHEP {\bf 0503}, 001 (2005).
\bibitem{SMITH}
  Y.~Hatta, E.~Iancu, L.~McLerran, A.~Stasto and D.~N.~Triantafyllopoulos,
  arXiv:hep-ph/0504182.

\bibitem{Balitsky05}
  I.~Balitsky,
  arXiv:hep-ph/0507237.
                                  
\bibitem{kl5}
  A.~Kovner and M.~Lublinsky, Nucl.Phys.A767:171-188, 2006,
  arXiv:hep-ph/0510047.

\bibitem{LL1}
E.~Levin and M.~Lublinsky,
Nucl.\ Phys.\, {\bf A730} (2004) 191.
\bibitem{LL2}
E.~Levin and M.~Lublinsky, Phys.\ Lett.\ {\bf B607}, (2005) 131.


\bibitem{first} J. Jalilian-Marian, A. Kovner, L. D. McLerran and H. Weigert, Phys.\ Rev.\ {\bf D55} (1997), 5414;
e-Print Archive: hep-ph/9606337;

\bibitem{second} Yu. V. Kovchegov, Phys.\ Rev.\ {\bf D54}:5463-5469,1996
e-Print Archive: hep-ph/9605446;

\bibitem{zakopane} A. Kovner, Lectures given at 45th Cracow School of Theoretical Physics, Zakopane, Tatra Mountains, Poland, 3-12 Jun 2005,  Acta Phys.Polon.B36:3551-3592,2005, e-Print Archive hep-ph/0508232;

\bibitem{bkw}R. Baier, A. Kovner, M. Nardi and U. A. Wiedemann, Phys.Rev.D72:094013,2005
e-Print Archive: hep-ph/0506126.

\bibitem{inprep} A. Kovner, M. Lublinsky and U. Wiedemann, in progress.

\bibitem{oderon} Y. Hatta, E. Iancu, K. Itakura and L. McLerran,  Nucl.Phys.A760:172-207,2005,
e-Print Archive: hep-ph/0501171.
 

\bibitem{genyakozlov1}  M. Kozlov and E. Levin, e-Print Archive: hep-ph/0604039.



\bibitem{genyakozlov} M. Kozlov and E. Levin, {\it Nucl.Phys.} {\bf A739} 291 (2004) e-Print Archive: hep-ph/0504146.


\bibitem{levintuchin} E. Levin and K. Tuchin,  Nucl.Phys.A693:787-798,2001
e-Print Archive: hep-ph/0101275; Nucl.Phys.A691:779-790,2001
e-Print Archive: hep-ph/0012167;  Nucl.Phys.B573:833-852,2000
e-Print Archive: hep-ph/9908317.


\end{thebibliography}
\end{document}